\documentclass[11pt]{article}

\usepackage{amsmath,amssymb}
\usepackage{txfonts}
\usepackage{graphicx}

\makeatletter
\long\def\@makecaption#1#2{%
  \vskip\abovecaptionskip
  \sbox\@tempboxa{#1. #2}%
  \ifdim \wd\@tempboxa >\hsize
    #1. #2\par
  \else
    \global \@minipagefalse
    \hb@xt@\hsize{\hfil\box\@tempboxa\hfil}%
  \fi
  \vskip\belowcaptionskip}

\def\@begintheorem#1#2{\trivlist 
   \item[\hskip \labelsep{\bfseries #1\ #2.}]\itshape}
\def\@opargbegintheorem#1#2#3{\trivlist
      \item[\hskip \labelsep{\bfseries #1\ #2\ (#3).}]\itshape}
\def\@endtheorem{\endtrivlist}

\def\@seccntformat#1{\csname the#1\endcsname.\ \ }
\makeatother

\title{Voronoi Diagrams for Pure 1-qubit Quantum States}
\author{Kimikazu Kato\thanks{Nihon Unisys / Department of Computer Science, University of Tokyo}, \
Mayumi Oto\thanks{COE Superrobust Computation Project, University of Tokyo}, \
Hiroshi Imai\thanks{ERATO Quantum Computation and Information / Department of Computer Science, University of Tokyo} \
and Keiko Imai\thanks{Department of Information and System Engineering, Chuo University}}
\date{}

\def\Tr{{\rm Tr}\;}
\newtheorem{theorem}{Theorem}
\newtheorem{lemma}{Lemma}

\begin{document}
\maketitle
\begin{abstract}

1-qubit quantum states form a space called the three-dimensional Bloch
ball. To compute Holevo capacity, Voronoi diagrams in the Bloch ball
with respect to the quantum divergence have been used as a powerful
tool.  These diagrams basically treat mixed quantum states corresponding
to points in the interior of the Bloch ball.  Due to the existence of
logarithm in the quantum divergence, the diagrams are not defined on
pure quantum states corresponding to points on the two-dimensional
sphere.  This paper first defines the Voronoi diagrams for pure quantum
states on the Bloch sphere by the Fubini-Study distance and the Bures
distance.  We also introduce other Voronoi diagrams on the sphere
obtained by taking a limit of Voronoi diagrams for mixed quantum states
by the quantum divergences in the Bloch ball.  These diagrams are shown
to be equivalent to the ordinary Voronoi diagram on the sphere.

\end{abstract}

\section{Introduction}

Quantum information has been attracting computer scientists as a new
computing paradigm \cite{nielsen-chuang}.  To develop a sound theory for
handling such quantum information, we need to understand the structure
of quantum information from the viewpoint of information processing.
Some aspect of quantum information is to define a kind of distance
between two quantum states.  Depending on the specific applications in
quantum estimation \cite{hayashi-ed}, quantum information geometry
\cite{amari-nagaoka}, quantum channel capacity \cite{holevo}, etc.,
there are many quantum distances, each having meanings in some
respective settings.

Voronoi diagrams have been playing a central role to represent the
proximity relation of point set, etc., with a wide variety of
applications in many fields.  It is quite natural to investigate the
proximity relation via Voronoi diagrams of quantum states with respect
to such distances.

Using the Voronoi diagrams of mixed quantum states with respect to the
quantum divergence, Oto, Imai and Imai \cite{oto} introduced a method to
calculate Holevo capacity.  Since there are many kinds of distances
between quantum states, we expect that a similar method can be
applicable to investigations of other distances.  In addition, to
investigate the difference between some two different distances of
quantum states, to see if their Voronoi diagrams coincide or not can be
a good first approach.  However, the Voronoi diagrams used in \cite{oto}
are not defined on pure quantum states. Therefore, it is reasonable to
investigate if the Voronoi diagrams used there can be extended to pure
states.  In 1-qubit case, pure states correspond to the surface of Bloch
ball, and mixed states to the interior. Geometrically, the problem can
be express as ``Can the Voronoi diagram defined only in the interior of
the Bloch ball be extended to its surface?''

Moreover, even from other points of view, pure quantum states are quite
important and useful. For example, most quantum algorithms have been
described using only pure states without using mixed states even if some
measurements are performed during the algorithm and quantum states are
then better to be treated as mixed ones.

In this paper, we first explain the method to calculate Holevo
 capacity. This method uses Voronoi diagrams for mixed quantum
 states by the quantum divergence. 
Secondly, we define Voronoi diagrams
for pure quantum states on the Bloch sphere by the Fubini-Study
distance and the Bures distance.  These diagrams are shown to be
equivalent to the ordinary Voronoi diagram on the sphere.  We also
introduce other Voronoi diagrams on the sphere which are obtained by
taking a limit of the Voronoi diagrams used in the calculation of
 Holevo capacity. Finally, all these diagrams --- the one by the Fubini-Study
 distance, the one by the Bures distance, and the one obtained by taking a
 limit of the diagram in mixed states --- are shown to be identical.

Consequently, as far as the proximity relation among 1-qubit pure states
are uniquely defined for these distances and divergences, it may be
regarded to be natural in a sense that geometric structures of pure
states are nicer than those of mixed states.

\section{Preliminaries}

\subsection{Bures distance and Fubini-Study distance}

A 1-qubit quantum state is represented by a density matrix $\rho$:
\[
\rho=\frac{1}{2}\left(
\begin{array}{cc}
1+z & x-{\rm i}y\\
x+{\rm i}y & 1-z
\end{array}
\right), \qquad x^2+y^2+z^2\leq 1 .
\]
$\rho=\rho(x,y,z)$ may be identified with a point $(x,y,z)$ in the
3-dimensional space, and a ball formed by all such points
\[
B=\{\,(x,y,z)\mid x^2+y^2+z^2\leq 1\}
\]
is called Bloch ball.  A state with rank 1 is called pure, while
a state with rank 2 is called mixed.  Points on the boundary of the
Bloch ball, i.e., the Bloch sphere, corresponds to pure states.

For two pure states $\rho$ and $\sigma$, the Fubini-Study distance 
$d_{\rm FS}(\rho,\sigma)$ is defined by
\[
\cos d_{\rm FS}(\rho,\sigma)=\sqrt{\Tr(\rho\sigma)},
\qquad 0\leq d_{\rm FS}(\rho,\sigma)\leq\frac{\pi}{2}.
\]
See Hayashi \cite{hayashi}.  The Bures distance $d_{\rm B}(\rho,\sigma)$
\cite{bures} is defined by
\[
d_{\rm B}(\rho,\sigma)=\sqrt{1-\Tr(\rho\sigma)}.
\]

\subsection{Quantum divergence and Holevo capacity}

Eigenvalues $\lambda_1,\lambda_2$ of $\rho(x,y,z)$ are given by
$(1\pm\sqrt{x^2+y^2+z^2})/2$.
By the eigenvalue decomposition, $\rho$ can be expressed as
$
\rho=\sum_i\lambda_i E_i,
$ where $E_i E_j$ is $E_i$ for $i=j$ and is 0 for $i\not=j$.
Then, for a mixed state $\rho(x,y,z)$,
$\log\rho$ is defined by
$
\log \rho=\sum_i(\log\lambda_i) E_i.
$
In the Bloch ball, information-geometric
structure can be induced by the von Neumann entropy $S(\rho)$
and the quantum divergence $D(\rho||\sigma)$.
The von Neumann entropy of a state $\rho$ is defined by
$
S(\rho)=\Tr(-\rho\log\rho).
$
Using the eigenvalues $\lambda_1,\lambda_2$ of $\rho$, it is
expressed as
$
S(\rho)=-\sum_i\lambda_i\log\lambda_i
$
i.e., $S(\rho)$ is the Shannon entropy of eigenvalues.
Note that $0\log 0=0$.
The quantum divergence $D(\rho||\sigma)$
for two quantum states $\rho$ and $\sigma$  is defined by
\[
D(\rho||\sigma)=\Tr(\rho(\log\rho-\log\sigma)).
\]
where $\sigma$ is a mixed state.
It is known that 
$D(\rho||\sigma)\ge0$, and $D(\rho||\sigma)=0$ iff $\rho=\sigma$.

Now we consider the situation of sending a qubit via a quantum channel
$\Gamma$ with noise and receiving it. A quantum channel means that
$\Gamma$ is a affine transformation that maps a quantum state to a
quantum state. If $\rho(x,y,z)$ is 1-qubit quantum state, the image of
$\Gamma$
\[
 \left\{ (x',y',z') \mid \rho'(x',y',z') =\Gamma(\rho(x,y,z)),
 (x,y,z)\in B \right\}
\]
is an ellipsoid and included in the Bloch ball.

The Holevo capacity of this quantum channel is known to be equal to
the maximum divergence from the center to a given point and the radius
of the smallest enclosing ball. The Holevo capacity $C(\Gamma)$ of a 1-qubit
quantum channel $\Gamma$ is defined as
\[
 C(\Gamma)= \inf_{\theta} \sup_{\rho} D(\Gamma(\rho)||\Gamma(\theta)),
\qquad (\theta,\rho\in B).
\]

\section{Computing the Holevo capacity of 1-qubit Quantum Channel}

The method to compute the Holevo capacity of a 1-qubit quantum channel
is described in \cite{oto}. There, the Voronoi diagrams of 1-qubit mixed
states by quantum divergence are used to solve the smallest enclosing
ball problem. The Voronoi diagrams were introduced as a generalization
of Kullback-Leibler divergence \cite{onishi-imai1,onishi-imai2}.  In
this section, we briefly explain the process of computation and its
mathematical background.

We plot sufficiently many points $V=\left\{ v_1,v_2, \ldots,v_N\right\}$
on the sphere and consider the Voronoi diagram of the points $\Gamma(V)$
with respect to the divergence. If $\# V$ is large enough, we can assume
that the radius of the smallest enclosing ball of the points $\Gamma(V)$
sufficiently approximates the real value of the Holevo capacity. To
compute the smallest enclosing ball, Voronoi diagrams are considered as
a useful tool.  Two Voronoi diagrams used here are defined as
\[
 V_D(v_i)=\bigcap_{i\neq j}\left\{\sigma | D(\sigma||\rho(v_i))\geq
 D(\sigma||\rho(v_j))\right\},
\]
\[ 
 V_{D^*}(v_i)=\bigcap_{i\neq j}\left\{\sigma | D(\rho(v_i)||\sigma)\geq D(\rho(v_j)||\sigma)\right\}.
\]

\begin{figure}
 \begin{minipage}[b]{0.5\hsize}
  \begin{center}
   \includegraphics[width=0.8\hsize,clip]{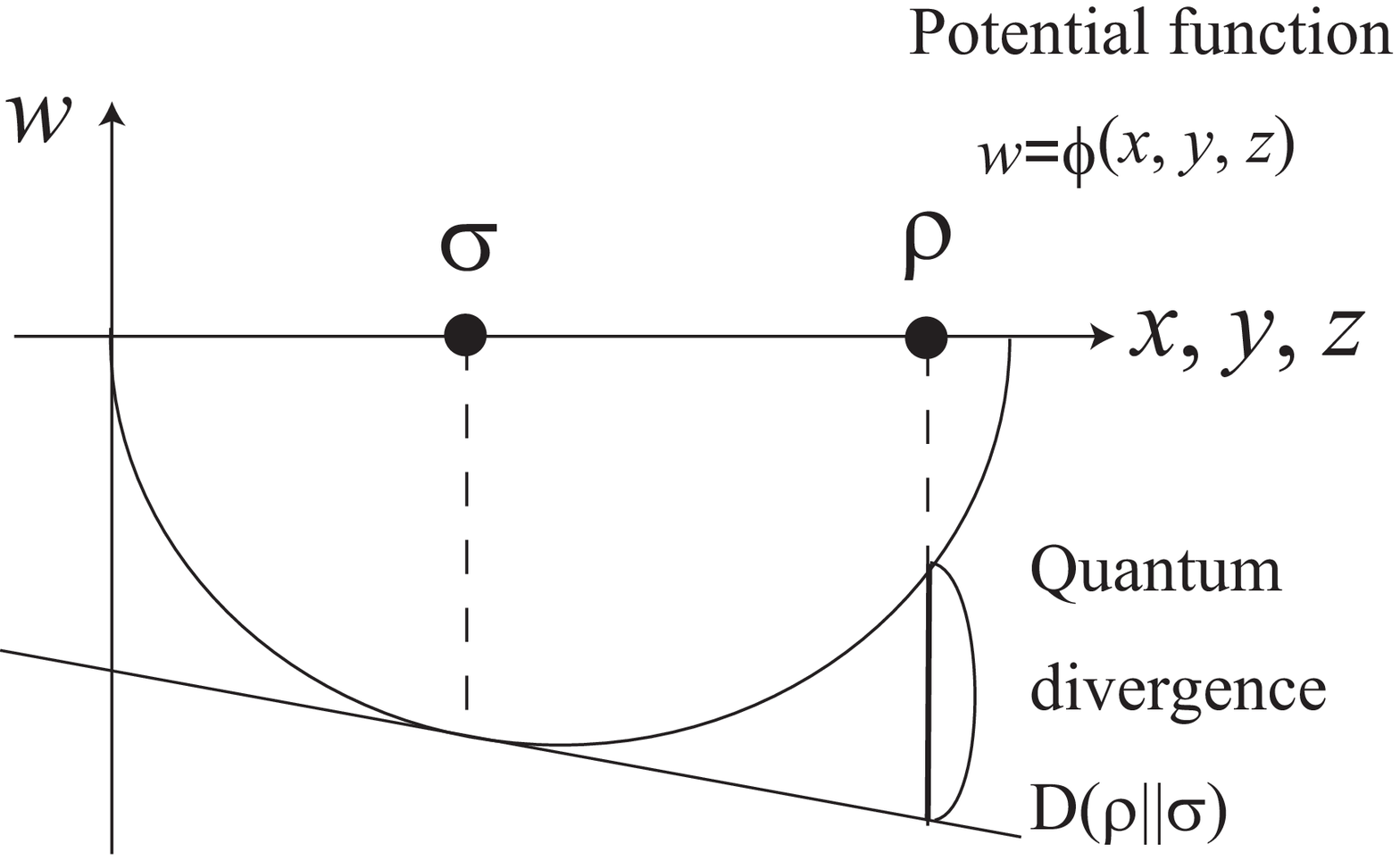}
   \caption{Geometric meaning of a divergence} \label{fig:distance-phi}
  \end{center}
 \end{minipage}
 \begin{minipage}[b]{0.5\hsize}
  \begin{center}
   \includegraphics[width=0.8\hsize,clip]{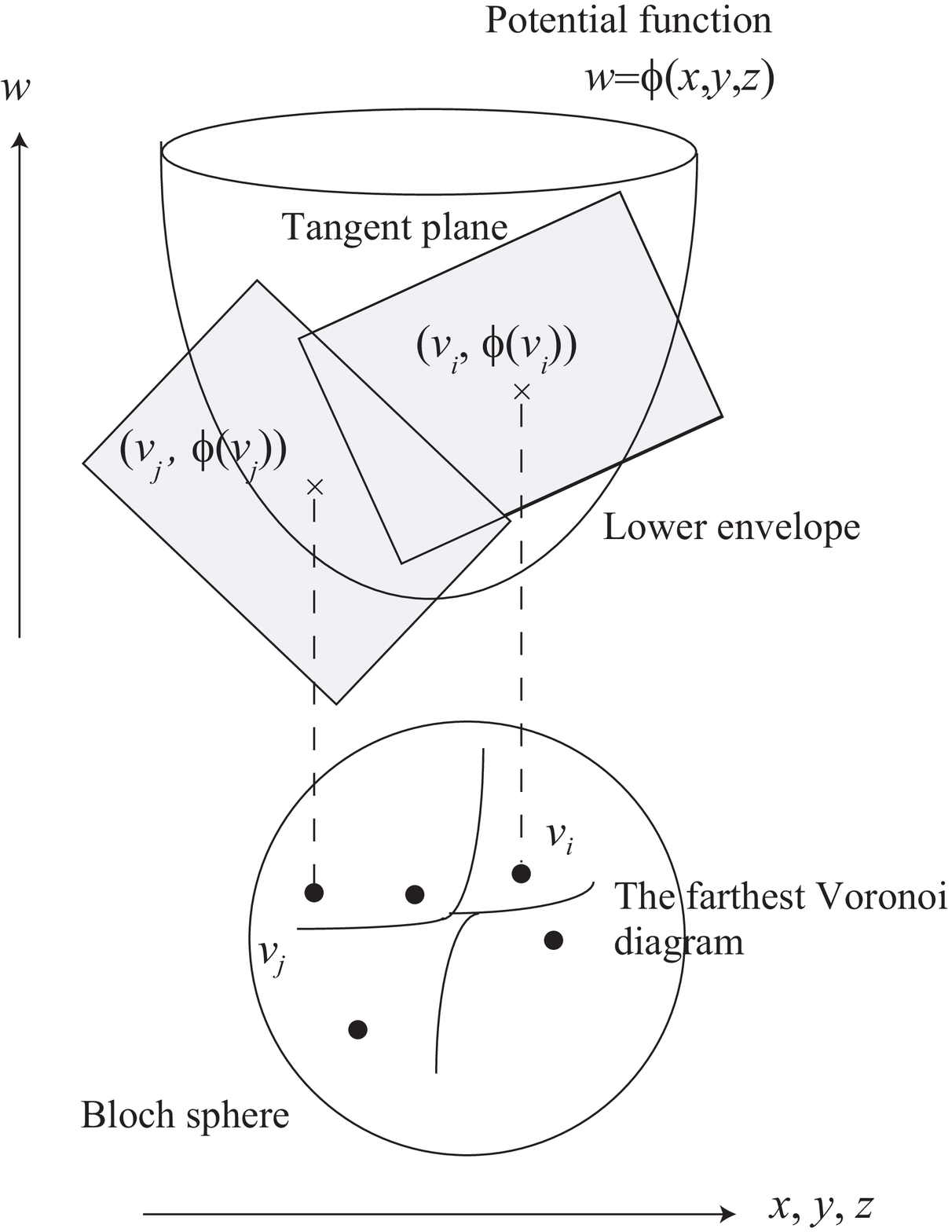}
   \caption{Computation of Voronoi diagram via a lower envelope}
   \label{fig:MES-phi}
  \end{center}
 \end{minipage}
\end{figure}

Although our concern is primarily $V_{D^*}$, we first consider $V_D$
because it is easier to compute. Actually, considering the graph of
$w=-S(\rho(x,y,z))=\phi(x,y,z)$,
$D(\rho(x,y,z)||\sigma(\tilde{x},\tilde{y},\tilde{z}))$ corresponds to
the distance along w-axis between the tangent plain at
$(\tilde{x},\tilde{y},\tilde{z})$ and the point $(x,y,z)$ (Fig.\
\ref{fig:distance-phi}). This fact makes it easy to compute the $V_D$
using a lower envelope of the tangent plains (Fig.\ \ref{fig:MES-phi}).

Then, we compute $V_{D^*}$ as a dual diagram of $V_D$.  To solve the
problem, we consider a dual coordinate system $\rho^*(u,v,w)$
corresponding to $\rho(x,y,z)$ such that
$D(\rho||\sigma)=D^*(\sigma^*||\rho^*)$. Their coordinate change is
explicitly defined as
\[
 (u,v,w)\equiv \nabla\phi = \frac{1}{2}\left(\log\frac{1+r}{1-r} \right)
\frac{1}{r}(x,y,z).
\] 
Thus, $V_{D^*}$ can be computed from $V_D$. In fact, this process to
work out $V_{D^*}$ from $V_D$ can be extended to a general
$d$-dimensional case \cite{oto}. Now using $V_{D^*}$, the center of the
smallest enclosing ball is determined and finally we obtain approximate
value of the Holevo capacity as the radius of the ball.

\section{Voronoi diagram of 1-qubit pure states}

We first consider the Fubini-Study distance $d_{\rm FS}(\rho,\sigma)$
for two 1-qubit pure states $\rho=\rho(x,y,z)$ and
$\sigma=\sigma(\tilde{x},\tilde{y},\tilde{z})$ with
$x^2+y^2+z^2=\tilde{x}^2+\tilde{y}^2+\tilde{z}^2=1$.  First, we have
\[
\Tr(\rho\sigma)
=\frac{1}{2}(1+x\tilde{x}+y\tilde{y}+z\tilde{z}).
\]
Hence, setting $\theta(\rho,\sigma)$ to be an angle between two vectors
$(x,y,z)$ and $(\tilde{x},\tilde{y},\tilde{z})$ with
$0\leq\theta\leq\pi$, we have
\[
d_{\rm FS}(\rho,\sigma)
=\cos^{-1}\sqrt{(1+\cos\theta(\rho,\sigma))/2}=\theta(\rho,\sigma)/2.
\]
In the 1-qubit case, the Fubini-Study distance between two pure states
is a half of the geodetic distance between two corresponding points on
the Bloch sphere.

Concerning the Bures distance,
\[
d_{\rm B}(\rho,\sigma)=
\sqrt{\frac{1}{2}(1-x\tilde{x}-y\tilde{y}-z\tilde{z}))}
=\frac{1}{\sqrt{2}}d_{\rm E}((x,y,z),(\tilde{x},\tilde{y},\tilde{z}))
\]
where $d_{\rm E}$ is the three-dimensional Euclidean distance.

Thus, at a first glance, these two distance might look strange, but in
the 1-qubit case both are just natural distances.  In fact, one we
restrict discussions on pure states only, these are more direct
consequences.  However, the above formulation is suitable to connect it
with mixed states.  This point will be discussed in the concluding
remarks.

Voronoi diagrams for pure 1-qubit states can be defined by using the
Fubini-Study distance or the Bures distance.  Suppose $n$ pure 1-qubit
states $\sigma_i=\sigma(x_i,y_i,z_i)$ $(i=1,\ldots,n)$ are given.
Define $V_{\rm FS}(\sigma_i)$ by
\[
V_{\rm FS}(\sigma_i)=\bigcap_{j\not=i}
\{(x,y,z)\mid \rho=\rho(x,y,z)\colon\ \hbox{pure states},
d_{\rm FS}(\rho,\sigma_i)<d_{\rm FS}(\rho,\sigma_j)
\}
\]
which is the Voronoi region of $\sigma_i$ with respect to the
Fubini-Study distance.  Similarly, $V_{\rm B}(\sigma_i)$ with respect to
the Bures distance can be defined.  Combining the above-mentioned
discussions with results on the ordinary Voronoi diagrams on the sphere
(e.g., \cite{imai-sumino-imai}), we have the following.

\begin{theorem}
\label{th1}
For $n$ 1-qubit pure states, the following four Voronoi
diagrams are equivalent:
\begin{enumerate}
\item
the Voronoi diagram with respect to the Fubini-Study distance
\item
the Voronoi diagram with respect to the Bures distance
\item
the Voronoi diagram on the sphere with respect to the ordinary
geodetic distance
\item
the section of the three-dimensional Euclidean Voronoi diagram with the
sphere
\end{enumerate}
\end{theorem} 

\section{Voronoi diagram of 1-qubit states by the quantum divergence}

As described above, the Voronoi diagram of 1-qubit states with respect
to the quantum divergence plays a very important roll in computation of
Holevo capacity. So far, the Voronoi diagrams are defined only in mixed
states.  Actually, while $D(\rho||\sigma)=\Tr\rho(\log\rho -
\log\sigma)$ can be defined when an eigenvalue of $\rho$ equals $0$
because $0\log 0$ can be naturally defined as $0$, it is not defined
when an eigenvalue of $\sigma$ is $0$. Here we show that this Voronoi
diagram of mixed states can be extended to pure states. In other words,
we prove that even though the divergence $D(\rho||\sigma)$ can not be
defined when $\sigma$ is a pure state, the Voronoi edges are naturally
extended to pure states.  To prove this convergence, we revisit the
geometric structure described in \cite{oto} by presenting explicit
expressions.

For a 1-qubit state $\rho=\rho(x,y,z)$ with 
$r\equiv\sqrt{x^2+y^2+z^2}$, the eigenvalues of $\rho$ is given by
\[
\lambda_1=\frac{1+r}{2},\qquad
\lambda_2=\frac{1-r}{2}.
\]
When $(x,y)\not=(0,0)$, defining a unitary matrix $U$ as
\[
U=\frac{1}{\sqrt{2}}
\left(
\begin{array}{cc}
 \displaystyle{\frac{x-{\rm i}y}{\sqrt{x^2+y^2}}\sqrt{\frac{r+z}{r}}} &
 \displaystyle{\frac{x-{\rm i}y}{\sqrt{x^2+y^2}}\sqrt{\frac{r-z}{r}}}
 \bigskip \\
 \displaystyle\sqrt{\frac{r-z}{r}} & -\displaystyle\sqrt{\frac{r+z}{r}} \\
\end{array}
\right),
\]
$\rho$ is expressed as
\[
\rho=U
\left(
 \begin{array}{cc}
  \lambda_1 & 0\\
  0 & \lambda_2\\
 \end{array}
\right)
U^*.
\]
Then,
\[
\begin{array}{rl}
\log\rho&={}
\displaystyle{
U
\left(
 \begin{array}{cc}
  \log\lambda_1 & 0\\
  0 & \log\lambda_2\\
 \end{array}
\right)
U^*
}\medskip\\
&={}
\displaystyle{
\frac{1}{2r}
\left(
\begin{array}{cc}
(r+z)\log\lambda_1+(r-z)\log\lambda_2 
& (x-{\rm i}y)(\log\lambda_1-\log\lambda_2)
\smallskip \\
(x+{\rm i}y))(\log\lambda_1-\log\lambda_2) 
& (r-z)\log\lambda_1+(r+z)\log\lambda_2 \\
\end{array}
\right)
}.
\end{array}
\]

For $\rho=\rho(x,y,z)$ and $\sigma=\sigma(\tilde{x},\tilde{y},\tilde{z})$
with $\tilde{r}=\sqrt{\tilde{x}^2+\tilde{y}^2+\tilde{z}^2}<1$,
$\tilde{\lambda}_1=(1+\tilde{r})/2$, and 
$\tilde{\lambda}_2=(1-\tilde{r})/2$,
we have the following:
\[
\arraycolsep=0pt
\begin{array}{rcl}
\Tr(\rho\log\sigma)&{}={}&
\displaystyle\frac{1}{4\tilde{r}}
( \log\tilde{\lambda}_1[(1+z)(\tilde{r}+\tilde{z})+2(x\tilde{x}+y\tilde{y})
  +(1-z)(\tilde{r}-\tilde{z})]\\
&&\qquad +
  \log\tilde{\lambda}_2[(1+z)(\tilde{r}-\tilde{z})-2(x\tilde{x}+y\tilde{y})
  +(1-z)(\tilde{r}+\tilde{z})]\\
&=&
 \displaystyle\frac{1}{2\tilde{r}}
 (\log\tilde{\lambda}_1[\tilde{r}+x\tilde{x}+y\tilde{y}+z\tilde{z}]
 +\log\tilde{\lambda}_2[\tilde{r}-x\tilde{x}-y\tilde{y}-z\tilde{z}])
 \smallskip\\
&=&\displaystyle{
 \frac{1}{2}\log\frac{1-\tilde{r}^2}{4}
 +\frac{\log(1+\tilde{r})-\log(1-\tilde{r})}{2\tilde{r}}
  [x\tilde{x}+y\tilde{y}+z\tilde{z}]
 }.\\
\end{array}
\]
When $\tilde{x}=\tilde{y}=0$ and $\tilde{z}\not=0$
for this $\sigma$, we have
\[
\Tr(\rho\log\sigma)=
 \frac{1+z}{2}\log\frac{1+\tilde{z}}{2}
+\frac{1-z}{2}\log\frac{1-\tilde{z}}{2}
=
 \frac{1}{2}\log\frac{1-\tilde{r}^2}{4}
 +\frac{\log(1+\tilde{r})-\log(1-\tilde{r})}{2\tilde{r}}
  z\tilde{z}
\]
and we now have the following.

\begin{lemma}
For a 1-qubit mixed state $\sigma=\sigma(\tilde{x},\tilde{y},\tilde{z})$
with $(\tilde{x},\tilde{y},\tilde{z})\not=(0,0,0)$
and a general 1-qubit state $\rho=\rho(x,y,z)$,
\[
D(\rho\|\sigma)=
\Tr(\rho\log\rho)-
 \frac{1}{2}\log\frac{1-\tilde{r}^2}{4}
 -\frac{\log(1+\tilde{r})-\log(1-\tilde{r})}{2\tilde{r}}
  [x\tilde{x}+y\tilde{y}+z\tilde{z}].
\]
Moreover, $D(\rho||\sigma)$ converges to $\Tr(\rho \log
 \rho)-1/2\log 1/4$ as $(x,y,z)\rightarrow(0,0,0)$.
This generalize the result and we can naturally assume that the above
 formula holds for all 1-qubit mixed states.
\end{lemma}

Now we revisit the Voronoi diagram defined by the quantum divergence.
For a set of mixed states $\sigma'_i=\sigma'_i(x'_i,y'_i,z'_i)$
$(i=1,\ldots,n)$,
we can define the Voronoi region $V_{D}(\rho_i)$ of $\sigma_i$.
Note that here $\sigma'_i$ should be mixed, while $\rho$ can be pure.
Suppose a set of pure states $\sigma_i=\sigma_i(x_i,y_i,z_i)$
$(i=1,\ldots,n)$ is given.
For a small $\epsilon>0$, consider the section of the Voronoi
diagram of $\sigma'_i(x'_i,y'_i,z'_i)$ defined by
\[
(x'_i,y'_i,z'_i)=(1-\epsilon)(x_i,y_i,z_i)
\]
with a sphere of $x^2+y^2+z^2=(1-\epsilon)^2$.  Define the Voronoi
diagram $V_{D}$ of these given pure states for all pure states by the
quantum divergence to be the limit of this section with
$\epsilon\rightarrow 0$, and denote the Voronoi region of $\sigma_i$
in this diagram by $V_{D}^{\rm pure}(\sigma_i)$.

Then, by using the above lemma, we have
\[
V_{D}^{\rm pure}(\sigma_i)
=\bigcap_{j\not=i}
 \{(x,y,z)\mid \rho=\rho(x,y,z)\hbox{ on the Bloch sphere}, \ 
x\tilde{x}_i+y\tilde{y}_i+z\tilde{z}_i
\ge 
x\tilde{x}_j+y\tilde{y}_j+z\tilde{z}_j
\}
\]
and see that this diagram is identical with those in the previous section.

Similarly as above, we can define $V_{D^*}(\sigma_i)$ for pure
$\sigma_i$, where $\rho$ should be mixed.  For the Voronoi diagram with
respect to the dual divergence $D^*$, consider its section with a sphere
of $x^2+y^2+z^2=(1-\epsilon)^2$, and define the Voronoi diagram
$V_{D^*}^{\rm pure}$ of these given pure states to be the limit of this
section with $\epsilon\rightarrow 0$.  The same arguments with
$V_{D}^{\rm pure}$ can be applied to this case, and we obtain the
following.

\begin{theorem}
The Voronoi diagrams $V_{D}^{\rm pure}$ and $V_{D^*}^{\rm pure}$
for $n$ pure states on the Bloch sphere are
identical to those diagrams in Theorem~\ref{th1}.
\end{theorem}

It should be noted that, for mixed states, the Voronoi diagrams $V_D$
and $V_{D^*}$ for the same set of quantum states are not identical in
general \cite{oto}.

\section{Concluding Remarks}

We have shown that in the 1-qubit case the Voronoi diagrams defined by
various distances and divergences of a finite set of pure states for
all pure states are all identical.  This would also hold in the
higher-dimensional case, which is left as an open problem.

Our investigations on pure states shed light on studying differences
among Voronoi diagrams with respect to many distances and divergences.
In fact, for pure quantum states, the Fubini-Study distance is a unique
metric once an appropriate differential-geometric invariance is imposed,
whereas for mixed quantum states there are so many metrics, such as SLD
Fisher metric, RLD Fisher metric and Bogoljubov Fisher metric (e.g., see
\cite{hayashi-ed}), each having some meaningful implications in some
settings.  In the case of fundamental information theory and statistics,
some relations between the Voronoi diagram by the Kullback-Leibler
divergence and that by the hyperbolic distance on the upper half-plane
are touched upon \cite{onishi-imai1,onishi-imai2,onishi-takayama}.
Investigating proximity relations induced by such metrics is also left
as a future work.

\end{document}